\documentclass[prb,twocolumn,superscriptaddress,noshowpacs]{revtex4}
\usepackage{graphicx}
\usepackage{physics}
\usepackage{bm}
\usepackage[utf8]{inputenc}

\begin{document}
\title{Analogue time machine in a photonic system}

\author{D.~D.~Solnyshkov}
\affiliation{Institut Pascal, PHOTON-N2, Universit\'e Clermont Auvergne, CNRS, SIGMA Clermont, F-63000 Clermont-Ferrand, France.}
\affiliation{Institut Universitaire de France (IUF), F-75231 Paris, France}

\author{G.~Malpuech}
\affiliation{Institut Pascal, PHOTON-N2, Universit\'e Clermont Auvergne, CNRS, SIGMA Clermont, F-63000 Clermont-Ferrand, France.}

\begin{abstract}
Analogue physics has successfully tackled the problems of chromodynamics, event horizons, Big Bang and Universe expansion, and many others. Here, we suggest a photonic model system for a "time machine" based on the paraxial beam approximation. We demonstrate how the closed time-like curves and the well-known grandfather paradox can be studied experimentally in this system. We show how the Novikov's self-consistency principle is realized in quantum mechanics thanks to the Heisenberg's uncertainty principle.
\end{abstract}
\maketitle

\section{Introduction}

Analogue physics is based on the mathematical similarities between different physical systems. These analogies often help to solve long-standing problems in some fields, by bringing the solutions known in the other fields. Among the most well-known examples of the success of such analogies are the Maxwell's equations (derived by analogy with the fluid dynamics in the presence of vortices) \cite{Maxwell1861}, the Anderson's suggestion \cite{Anderson1963} for the mass generation for bosons by symmetry breaking via what is now known as Higgs mechanism \cite{Higgs1964} (based on the analogy with superconductors), and the Semenoff's proposal for the realization of the Dirac's Hamiltonian in graphene \cite{Semenoff1984} with the associated effect of Klein tunneling \cite{Klein1929}, which stimulated the works on graphene \cite{katsnelson2006chiral}. One of the particularly developed fields of analogue physics is analogue gravity \cite{Barcelo2005,Barcelo2018}. Thirty years after the initial proposal of Unruh \cite{Unruh1981} for the observation of an analogue of the Hawking emission \cite{Hawking1974} expected to arise at event horizons, such emission has indeed been observed experimentally in classical \cite{Weinfurtner2011} and quantum fluids \cite{Steinhauer2014,Steinhauer2019}. The studies of the analogue spacetimes are not limited to the Hawking emission:  they include also the superradiance \cite{Torres2017} and the Penrose effect \cite{Solnyshkov2019} for the Kerr black holes, the Big Bang and the expansion \cite{Eckel2018}, analogue wormholes \cite{Solnyshkov2011}, and even the false vacuum decay \cite{braden2019nonlinear,Billam2019}.

A very interesting conclusion of the general relativity is the possibility of existence of closed time-like curves (CTCs), commonly called "time machines" \cite{Morris1988}.
Indeed, the relativity of simultaneity implies that there is no common "now" for the whole Universe, and therefore that different moments of time coexist, and can possibly be traveled to. Soon after the discovery of the wormhole-type solutions of the Einstein's equations for the spacetime metric \cite{Einstein1935,Fuller1962}, it was understood that traversable wormholes \cite{Ellis1973,Bronnikov1973} allowed faster-than-light travel, which, in turn, makes possible the time travel \cite{Morris1988}. Moreover, it seems that any faster-then-light travel, which is a goal of current NASA projects \cite{White2013}, appears as backward time motion for some observers. The CTCs are the spacetime trajectories of the objects traveling through a time machine: their closed character implies the possibility for the object to affect its own past. The most well-known theoretical result concerning the CTCs is the Novikov's self-consistency principle stating that the only events which can occur along such closed curves are those which are globally self-consistent \cite{Friedman1990}. These works have inspired a strong research activity on both classical \cite{Echeverria1991} and quantum \cite{Deutsch1991} problems in presence of time machines, including the problem of free will \cite{Tobar2020}. In quantum mechanics with CTCs, many counter-intuitive results were obtained, such as the solution of NP-complete problems in polynomial time \cite{Bacon2004}. Recently, the interaction of a qubit with another one trapped in a CTC has been simulated experimentally, marking another milestone for analogue physics \cite{Ringbauer2014}. The long-standing question of casualty, free will, and the possibility of changing one's own past, has moved from the realm of philosophical problems into the dominion of experimental physics.

Photonics offers extended possibilities for analogue physics. The whole field of topological photonics \cite{lu2014topological,Ozawa2019} was born from the possibility to simulate wavefunctions in periodic crystal lattices using electromagnetic waves in artificially constructed periodic media. The advantage of photonics is in the opportunity to observe the wavefunctions experimentally (including the phase) and to perform the wavefunction engineering with artificial potentials, e.g. periodic lattices. As an example, the possibilities to electromagnetic wormholes with metamaterials were recently suggested \cite{greenleaf2007electromagnetic,prat2015magnetic}. In particular, the well-known paraxial approximation for light has been used for analogue physics studies in atomic vapor cells \cite{Zhang2016,Zhang2019}, nonlinear crystals \cite{Vocke2018}, and in coupled waveguides \cite{rechtsman2013photonic}. 

In this work, we propose to study the self-consistency of time travel by using the equivalence between the time coordinate in the Schrodinger equation and one of the space coordinates (the $z$ coordinate, corresponding to the beam propagation direction) in the paraxial configuration. 
We show with numerical simulations that the system indeed converges to a stationary self-consistent solution, confirming the Novikov's principle, which is realized thanks to the inherent quantum uncertainty.  We demonstrate that in the stationary case, the time-looped signal can be either self-amplified or self-suppressed. We show that this suppression can never be complete. Traveling to the past and killing the younger version of yourself is therefore impossible. 

\section{The model}
The paraxial approximation consists in considering the envelope of the electric field and neglecting $\partial^2 \bm{E}_{\perp}/\partial z^2$ with respect to $k_0\partial\bm{E}_{\perp}/\partial z$  in the Laplacian term of the Helmholtz equation for the electromagnetic field. The resulting equation for the envelope (neglecting the spin-orbit coupling effects) reads
\begin{equation}
    i\frac{\partial E}{\partial z}=-\frac{1}{2 k_0}\left(\frac{\partial^2}{\partial x^2}+\frac{\partial^2}{\partial y^2}\right)E-\frac{k_0\chi}{2} E
    \label{parax}
\end{equation}
where $\chi$ is the dielectric susceptibility of the medium and $k_0$ is the propagation wave vector along $z$. This equation is equivalent to the time-dependent 2D Schr\"odinger equation
\begin{equation}
    i\hbar\frac{\partial\psi}{\partial t}=-\frac{\hbar^2}{2m}\Delta\psi +U\psi
    \label{tSchr}
\end{equation}
with the mass $m$ determined by $k_0$ and the potential $U$ determined by the susceptibility profile $\chi$. Nonlinear terms can be present in both equations. Their role will be considered in the following sections.

\begin{figure}[tbp]
    \centering
    \includegraphics[width=0.95\linewidth]{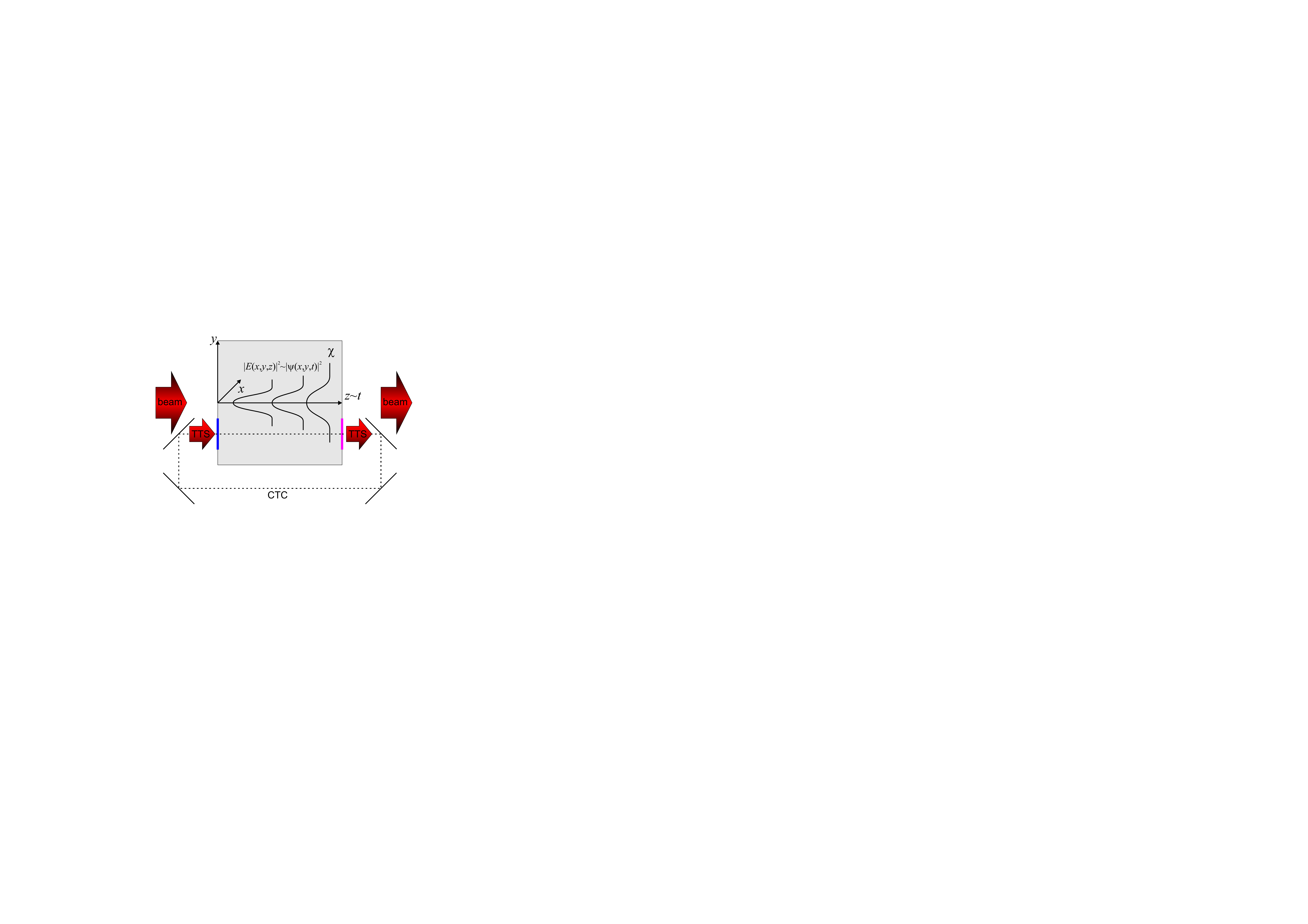}
    \caption{Scheme of the experiment. The dielectric medium characterized by susceptibility $\chi$ is shown as a gray rectangle. The input beam arrives from the left. The $z$-dependent envelope of the electric field $E(x,y,z)$ behaves as a time-dependent wavefunction $\psi(x,y,t)$, its diffusion is sketched with black lines.
    The output region marked with a magenta line is transferred to the input region marked with a blue line using 4 mirrors (Mach-Zehnder interferometer setup). The time-traveler signal (TTS) goes around the closed-time like curve (CTC) marked with a dashed line. }
    \label{fig1}
\end{figure}

The $z$ coordinate of Eq.~\eqref{parax} maps to the time variable in the Schr\"odinger equation~\eqref{tSchr}. We propose to use this mapping to create a model of a CTC or a time machine by coupling the output of the medium to its input as shown in Fig.~\ref{fig1}. This analogue system is not supposed to reproduce the wormhole itself, which is a relativistic object, but rather its effect on the asymptotically flat regions of spacetime. The initial beam from the source enters the  medium described by the Eqns.~\eqref{parax},\eqref{tSchr} from the left. The evolution of the electric field envelope (or wavefunction) with $z$ (analogue time) is sketched by a black solid line. A set of 4 mirrors (forming a Mach-Zehnder interferometer) loop a part of the beam from the output on the right back to the entrance on the left, forming the CTC (dashed line). Only the internal part of the CTC is simulated in the analogue system, whereas the external part (the wormhole) is supposed to transmit the signal with minimal changes (as expected for a time machine).

\section{Linear medium}

We begin by considering the problem of the possible stationary solution for the electric field envelope $E(x,y,z)$ equivalent to the existence of a stable history of the Universe $\psi(x,y,t)$ mathematically. Let us consider that the signal from a single point $x_0,y_0$ at the moment $T$ given by $\psi(x_0,y_0,T)$ is sent backwards into the past to the same point at the moment $t=0$ and adds to the wavefunction already existing at that moment as
\begin{equation}
\psi(x_0,y_0,0)\leftarrow\psi(x_0,y_0,0)+\alpha \psi(x_0,y_0,T)
\label{ttlin}
\end{equation}
where $\alpha$ is a complex coefficient describing the efficiency of the time machine. In a realistic optical system without gain $|\alpha|\le 1$, but in principle it can also exceed unity.
The most efficient way to find a stationary solution for $E$ is to use iterations: solve the time-dependent equation for $\psi$ from $t=0$ to $t=T$, and then use $\psi(x_0,y_0,T)$ as the input for the next iteration. We can therefore write  
\begin{equation}
    \psi_{n+1}(x_0,y_0,0)\to\psi_n(x_0,y_0,0)+\alpha \psi_n(x_0,y_0,T)
    \label{itern}
\end{equation}
From the mathematical point of view, the values of the signal sent backwards (describing the state of the time-traveler) represent a sequence. This sequence is either convergent or not. If it converges to a certain limit $\psi_n(x_0,y_0,T)\to c$, four interesting situations are possible at the first glance:\begin{enumerate}
    \item $c=0$, complete suppression of the time-traveler signal (TTS);
    \item $|c|<|\psi_0(x_0,y_0,T)|$, partial suppression of the TTS;
    \item $|c|>|\psi_0(x_0,y_0,T)|$, limited amplification of the TTS;
    \item $c=\infty$, unlimited amplification of the TTS.
\end{enumerate}

Let us first focus on the first possibility, which corresponds precisely to the time-traveler (or grandfather) paradox: is it possible to kill one's own younger self? The answer that we can prove mathematically is NO. Indeed, let us suppose that $c=\lim\psi_n(x_0,y_0,T)=0$. By definition, it means that there exists an $N$ such that for $n>N$, $|\psi_n(x_0,y_0,T)|<\epsilon$ (with arbitrary small $\epsilon$). Then, for the next iteration $n+1$ at $t=0$ we have an initial value which is also arbitrarily close to the \emph{zeroeth} (that is, initial) iteration: \begin{equation}
\left|\psi_{n+1}(x_0,y_0,0)-\psi_0(x_0,y_0,0)\right|<|\alpha|\epsilon,
\end{equation}
because this initial value is given by 
\begin{equation}
\psi_{n+1}(x_0,y_0,0)=\psi_{0}(x_0,y_0,0)+\alpha\psi_n(x_0,y_0,T)
\end{equation}
For any finite $|\alpha|$, the product $|\alpha|\epsilon$ can be made arbitrary small. Next, we apply the Lyapunov analysis of the stability \cite{Eckmann1985}. If the original system (without the CTC) is stable (not chaotic) and all Lyapunov exponents are negative, the arbitrary small separation of initial conditions (at $t=0$) implies even smaller separation of the final values (at the moment $T$):
\begin{widetext}
\begin{equation}
|\psi_{n+1}(x_0,y_0,T)-\psi_0(x_0,y_0,T)|<|\psi_{n+1}(x_0,y_0,0)-\psi_0(x_0,y_0,0)|<|\alpha|\epsilon.
\end{equation}
\end{widetext} 
We find therefore that $\psi_{n+1}$ is arbitrary close to zero and at the same time to the zeroth iteration value $\psi_0(x_0,y_0,T)$, which is impossible, unless $\psi_0(x_0,y_0,T)=0$ (there was no time traveler from the start, which is a trivial situation). We must conclude that the signal sent to the past cannot suppress itself completely.

The three other configurations are possible mathematically. We note, however, that the infinite self-amplification is impossible in a physical implementation of an analogue time machine that we suggest because of the inevitable gain saturation mechanisms. We conclude that in practice, the TTS is either partially suppressed or amplified.

It is also possible that the sequence $\psi_n$ does not converge at all: the stationary configuration of electromagnetic field does not settle down in the system. This can occur if the original system has a positive Lyapunov exponent (exhibits a chaotic behavior), because in this case even a weak signal into the past strongly modifies the future, including itself. The possibility of creation of such a configuration in an analogue system in the quantum case that requires dynamical quantum chaos \cite{pokharel2018chaos} remains an open question that we leave for future works. 

\begin{figure}[tbp]
    \centering
    \includegraphics[width=0.95\linewidth]{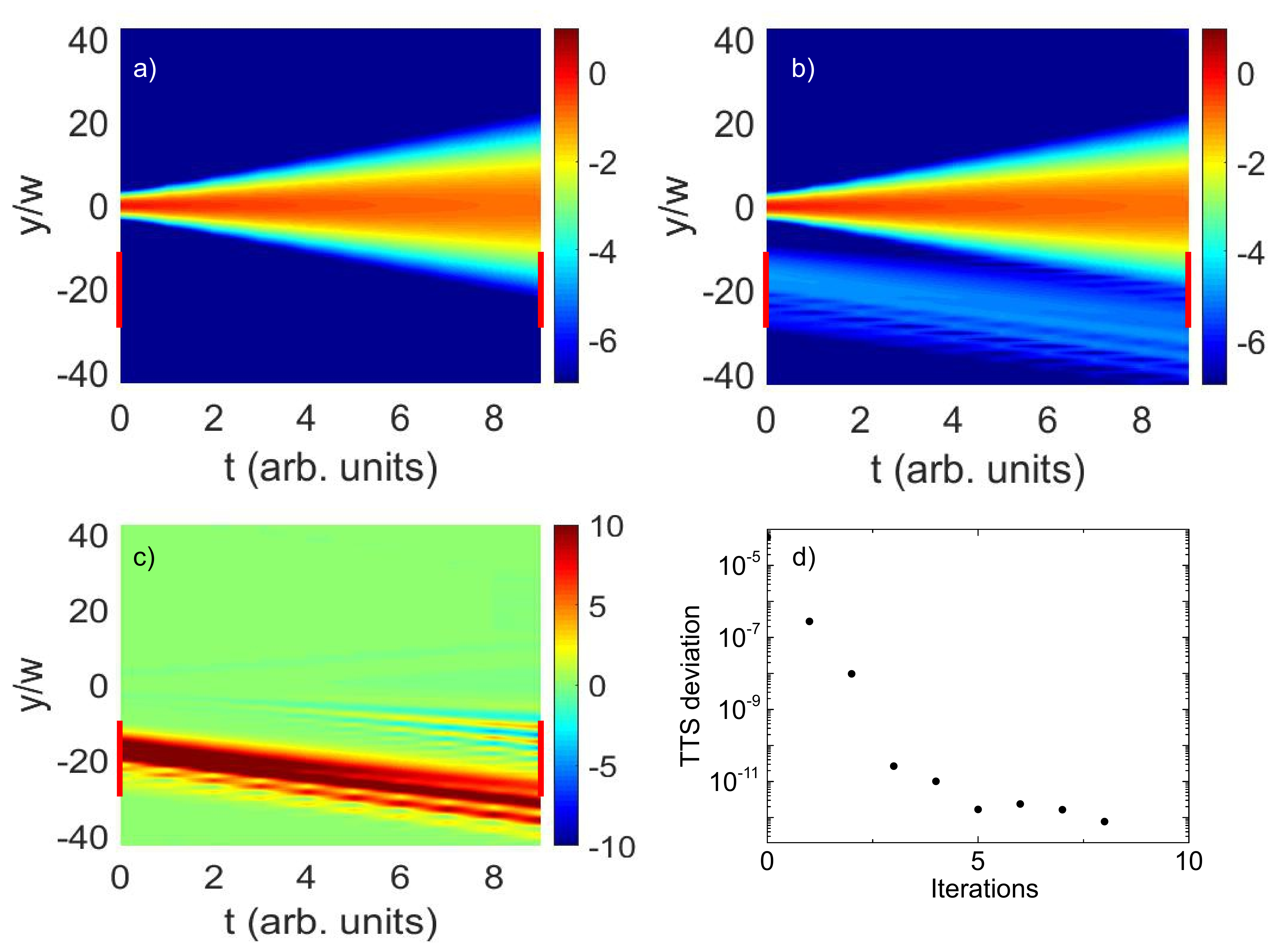}
    \caption{Numerical simulations. a) Initial distribution of $|\psi(y,t)|^2$ (without CTC), log scale; b) Final stationary distribution of $|\psi(y,t)|^2$ (with CTC), log scale; c) Difference between the initial and the stationary distributions, linear scale; d) convergence of the TTS integrated density with iteration number. Red lines mark the CTC windows.}
    \label{fig2}
\end{figure}

Figure~\ref{fig2} shows the results of numerical simulations based on the linear time-dependent Schr\"odinger equation with a Gaussian spatial profile of the CTC (describing the mirrors shown in Fig.~\ref{fig1}) coupling the final value of the wave function $\psi(y,T)$ into its initial value $\psi(y,0)$. The problem is reduced to 1D (the system is considered to be homogeneous along $x$) in order to focus on its essential features. The simulations confirm the possibility of the achievement of a stationary solution for the electromagnetic field, an equivalent of  a self-consistent history in presence of a CTC. Fig.~\ref{fig2}(a) shows the initial distribution of the probability density $|\psi(y,t)|^2$ (no CTC), while Fig.~\ref{fig2}(b) shows the stationary distribution. The TTS is clearly visible in the bottom part of the figure. It presents a non-zero wave vector oriented downwards, because only such components (present in the original beam) can penetrate into the CTC window located at around $y=-20$~$\mu$m. Fig.~\ref{fig2}(c) shows the difference of the probability density between the initial and the stationary configurations. The TTS appears as red (local probability increase), but blue regions are also visible: the TTS locally exhibits a negative interference with the initial beam. Finally, Fig.~\ref{fig2}(d) demonstrates the convergence of the system: the deviation from the final (stationary) solution drops down to the machine precision in just about $5$ iterations. The sequence $\psi_n(x_0,y_0,T)$ converges to a value about $11\%$ larger than the initial value $\psi_0(x_0,y_0,T)$.
This numerical simulation confirms that a stationary solution can be found in this case and that the system can exhibit a self-consistent history, at least in the linear case.

The fast convergence is ensured not only by the fact that this simplest quantum system is not chaotic, but even stronger by its diffractive nature. The spreading of the wave packets guarantees that $|\psi_n(x_0,y_0,T)|^2<|\psi_n(x_0,y_0,0)|^2$, which means that the TTS is weakened with each iteration, and thus the sequence rapidly converges. The non-relativistic quantum mechanics with a CTC appears to respect the self-consistency principle.

\begin{figure}[tbp]
    \centering
    \includegraphics[width=0.95\linewidth]{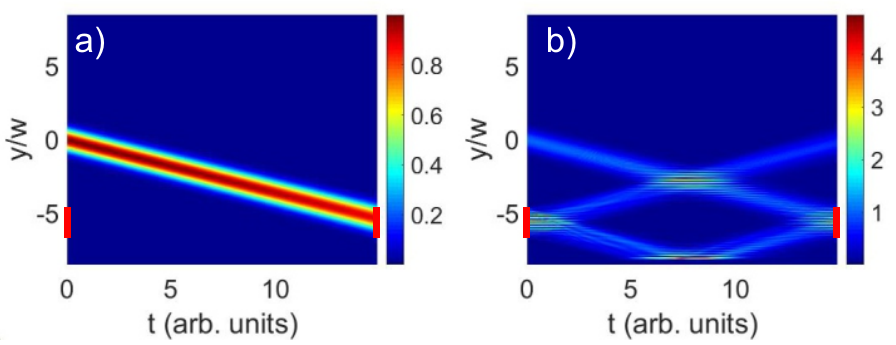}
    \caption{Numerical simulations showing a more complicated configuration with an initially propagating wavepacket. a) Initial distribution of $|\psi(y,t)|^2$ (without CTC), linear scale; b) Final stationary distribution of $|\psi(y,t)|^2$ (with CTC), linear scale. Red lines mark the CTC windows. }
    \label{fig3}
\end{figure}

A less trivial example is shown in Fig.~\ref{fig3}: the wavepacket has a non-zero initial wave vector along $y$ (Fig.~\ref{fig3}(a), $k_y<0$), directed towards the "time machine" (red line). It enters the time machine and reappears at $t=0$, still directed downwards. Then the wave packet is reflected at a potential barrier at the boundary of the system and gets a new wave vector $k_y>0$ directed upwards. It then reenters the CTC and reappears at $t=0$ once again, this time continuing to propagate upwards. \emph{Three} copies of the same wave packet are present in the system at any moment of time. This configuration is close to that of the classical billiard ball problems \cite{Friedman1990}, with a time-travelling billiard ball hitting a previous version of itself. One could expect that the wave packets might interfere destructively at some point, either at $t=0$ or at $t=T$, and thus suppress the TTS, but this does not happen, because of the non-zero wave vector: actually, an interference pattern between positive and negative $k_y$ is observed at $t=0$ and at $t=T$, and an extra phase simply shifts this interference upwards or downwards, but it does not lead to the complete suppression of the TTS. The convergence in this case is as fast as in the configuration of Fig.~\ref{fig2}.

\begin{figure}[tbp]
    \centering
    \includegraphics[width=0.95\linewidth]{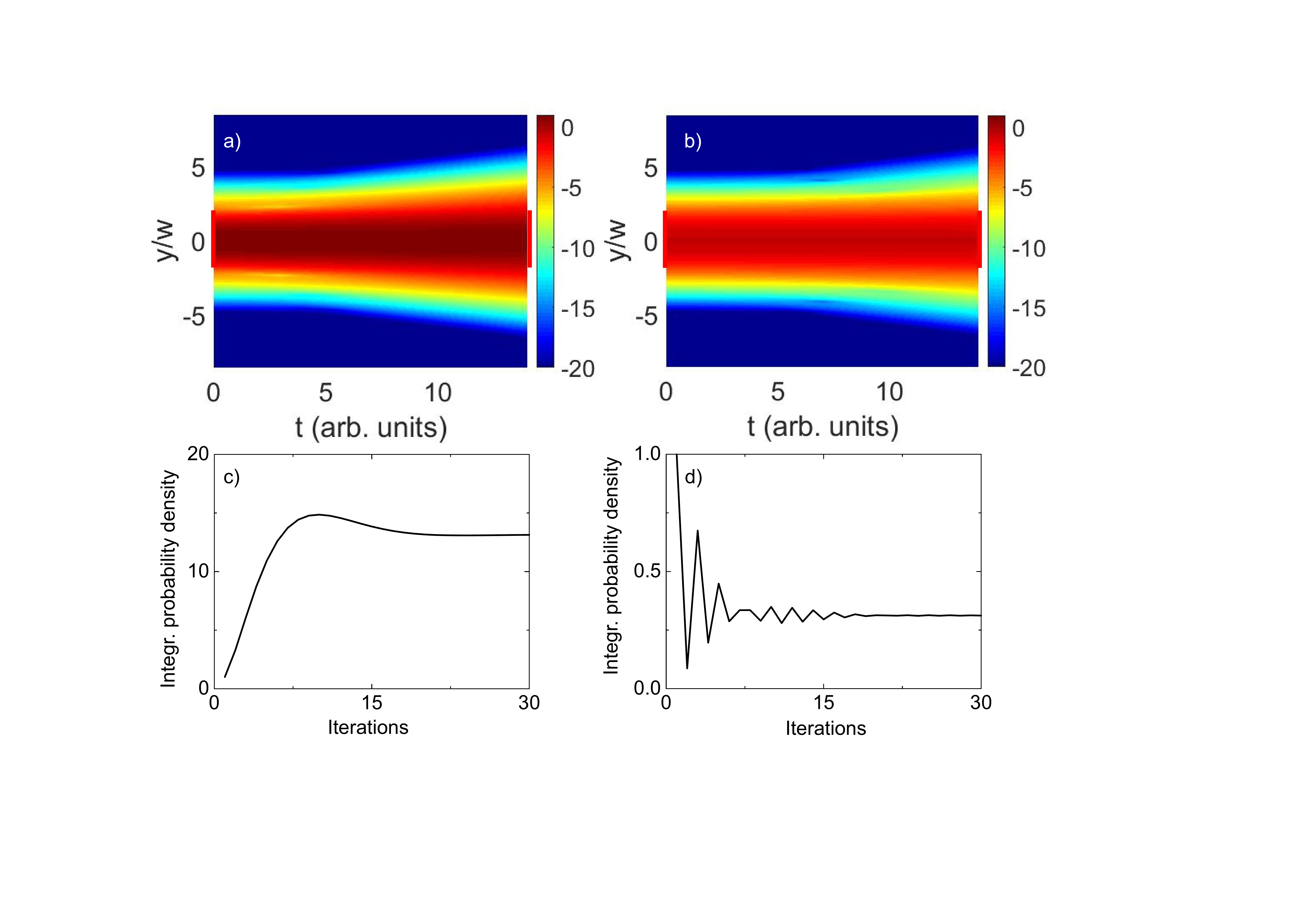}
    \caption{Numerical simulations of the strongest self-affecting configuration. (a,b) Final stationary distribution of $|\psi(y,t)|^2$ (with CTC), log scale. Coupling phase: (a) $\phi=0$, (b) $\phi=\pi$. (c,d) Corresponding total probability density exhibiting convergence.}
    \label{fig4}
\end{figure}

The most self-affecting configuration corresponds to a wide beam with $k_y=0$, exhibiting almost no diffraction and entering straight into the CTC, as shown in Fig.~\ref{fig4}. This beam reappears at $t=0$ with the same wave vector $k_y=0$ and thus reenters the CTC again and again. Two situations limiting cases are possible: the phase of the CTC coupling coefficient $\phi=\arg\alpha$ can be either zero (constructive interference, Fig.~\ref{fig4}(a,c)) or $\pi$ (destructive interference, Fig.~\ref{fig4}(b,d)). In the first case, we could expect that the wave packet might exhibit unlimited amplification and thus no stationary solution could be reached. This does not happen, as can be seen from panel (c), showing the integrated probability density, whose value saturates around $12$ (twelve copies of the wave packet simultaneously present in the system). Such cases are rarely treated in science fiction literature (but see e.g. Ref. \cite{LemStar}). Diffraction prevents further increase of the self-amplification. Similarly, diffraction prevents the complete suppression of the beam in panel (b) ($\phi=\pi$). The main part of the beam is strongly suppressed, the integrated probability initially drops down to $\approx 0.1$, meaning that the time traveler managed to kill its own former self by about 90\%. But the complete suppression does not occur, in agreement with our analytical predictions. The system converges in about 30 iterations to a value of about 30\% of the initial wave packet, exhibiting a strongly broadened spatial distribution. We can conclude that the Novikov's self-consistency principle is ensured in quantum mechanics by the Heisenberg's uncertainty principle.

\section{Non-linear medium}

The paraxial propagation of a beam in a non-linear medium is described by the non-linear Schrodinger equation:
\begin{equation}
    i\hbar\frac{\partial\psi}{\partial t}=-\frac{\hbar^2}{2m}\Delta\psi+g\left|\psi\right|^2\psi +U\psi
    \label{NLSE}
\end{equation}
 Depending on the sign of the non-linearity, the interactions (characterized by the constant $g$) can be either attractive ($g<0$, in which case the system is unstable) or repulsive ($g>0$). The evolution of long-wavelength weak excitations (characterized by a speed of sound $c_s=\sqrt{g n/m}$) in such system can be described using a relativistic wave equation \cite{Garay2000}:
\begin{equation}
\partial_{\nu}(\sqrt{-g}g^{\mu\nu}\partial_{\nu}\varphi)=0
\end{equation}
written with an effective metric tensor $g_{\mu\nu}$ totally determined \cite{Unruh1981} by the background stationary velocity $\textbf{v}=(\hbar/m)\nabla\arg\psi$ and the local speed of sound $c_s$: 
\begin{equation}
g_{\mu\nu}=\frac{mn}{c}\begin{pmatrix} 
-(c^{2}-\textbf{v}^{2}) & \vdots & -\textbf{v}\\
 \ldots \ldots \ldots & . & \ldots \ldots \\
-\textbf{v} & \vdots & \delta_{ij}\\
 \end{pmatrix} \label{gmetric}
\end{equation} 
It is already widely used for the analogue studies of general relativity, including time-related effects. This regime allows to map the electromagnetic wave in a non-linear crystal not to the non-relativistic quantum mechanics, as in the previous section, but to relativistic wave equation for light.
Moreover, this nonlinear system also admits another type of non-trivial solutions: solitons, localized density perturbations characterized by a phase jump and behaving as relativistic massive particles \cite{Pitaevskii}. These have also already been used for analogue studies \cite{Solnyshkov2012,Hivet}. 

We note that in this case the physical signal to be sent along the CTC in the analogue system should be taken as a deviation from the mean value of the electric field, and not the total electric field as in \eqref{ttlin}:
\begin{equation}
\psi(x_0,y_0,0)\leftarrow\psi(x_0,y_0,0)+\alpha \left(\psi(x_0,y_0,T)-\bar{\psi}(x,y,T)\right)
\label{ttnl}
\end{equation}
 This is best achieved in a pump-probe configuration, where the pump can be suppressed by interference.

This analogue system is much closer to the problems of general relativity and exhibits a much richer behavior than the linear system. In particular, it is possible to simulate the configuration of the billiard ball striking the former version of itself \cite{Echeverria1991}. Shallow solitons have the advantage of being relatively weakly interacting and propagating almost at the speed of sound. Their presence also does not significantly perturb the phase of the condensate.

\begin{figure}[tbp]
    \centering
    \includegraphics[width=0.95\linewidth]{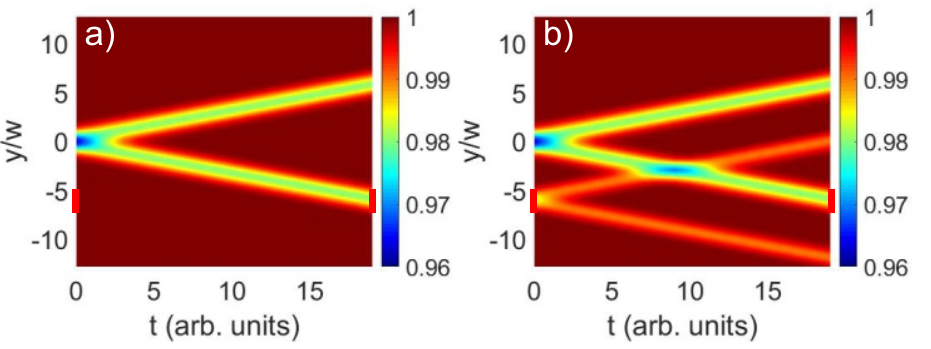}
    \caption{Numerical simulations for a nonlinear system. a) Initial distribution of $|\psi(y,t)|^2$ (without CTC), linear scale; b) Final stationary distribution of $|\psi(y,t)|^2$ (with CTC), linear scale. Red lines mark the CTC windows. }
    \label{fig5}
\end{figure}

Figure~\ref{fig5} shows the results of numerical simulations of shallow soliton propagation in a non-linear system. The initially created density minimum separates into two gray solitons propagating almost at the speed of sound in the medium $c_s$. One of them enters a CTC and reappears at $t=0$, generating two more shallow solitons, the trajectory of one of which crosses one of the initial solitons (panel (b)). At high speeds, the interaction between the solitons is relatively weak, and even though it leads to a slight deviation of the first soliton, it does not prevent it from entering the CTC. Thus, the time traveler does not destroy itself completely once again, and the system exhibits a self-consistent history.

The characteristic speed of the propagation of the changes in the history in this model system is fundamentally different from the analogue of the speed of light. Indeed, the role of the speed of light is played by $c_s$, whereas the changes propagate from the past to the future along the equivalent of the time axis $z$ with the speed of light in the medium $c/n$. It means that a change in the past affects the time traveler not immediately, for some moment (of the laboratory time) the traveler "remembers" the previous version of the history.

\section{Discussion}

If the past, the present, and the future exist simultaneously, it means the future "already" exists, and our free will arises only from the impossibility to predict the future (which is already existing anyway). In this case, the only possibility of a real change in the Universe is given by the time machines. The time travel changes the history and thus changes the past and the future. On the other hand, the final (stationary) version of a history in presence of a time machine is not so different from what we actually experience. We have seen that a signal from the future can change the history in a quite significant way, and the only certain point is that this signal cannot disappear completely as a result of these changes. But the history does not keep track of its initial version, only the final one can be observed.

Ultimately, any time machine can be represented as a system with a feedback in a stationary equilibrium configuration. The behavior of the systems with a feedback has been studied by the control theory in a lot of works. Our own brain is a complicated feedback system, and in this sense, works like a time machine: we conceive a certain future which is ultimately not realized, because we get some information from this potential future and adapt our behavior correspondingly, in order to optimize the outcomes. But the realized version of the history contains our "mind simulations" and the feedback signal that we have received from them as a part of our personal history.

\section{Conclusions}

To conclude, we have shown that it is possible to simulate "time machines" or systems with closed time-like curves, using electromagnetic beams in the paraxial configuration, mapped to time-dependent Sch\"odinger equation. We have shown that it is possible to check the Novikov's self-consistency principle experimentally in such systems. Our analysis demonstrates how the time-traveler (or the grandfather) paradox is resolved in quantum mechanics. We have shown that the self-consistency is ultimately achieved thanks to the Heisenberg's uncertainty principle. However, it might be violated in systems with dynamical quantum chaos.

\begin{acknowledgments}
We acknowledge the support of the projects EU "QUANTOPOL" (846353), "Quantum Fluids of Light"  (ANR-16-CE30-0021), of the ANR Labex GaNEXT (ANR-11-LABX-0014), and of the ANR program "Investissements d'Avenir" through the IDEX-ISITE initiative 16-IDEX-0001 (CAP 20-25). 
\end{acknowledgments}

\bibliography{biblio}

\end{document}